\documentclass[12pt]{article}
\usepackage{epsf}
\renewcommand{\i}{\mathrm{i}}       

\newcommand{\be}{\begin{eqnarray}}  
\newcommand{\ee}{\end{eqnarray}}    
\newcommand{\e}{\mathrm{e}}         
\renewcommand{\d}{\mathrm{d}\!}     
\newcommand{\D}{\mathrm{D}\!}       

\begin{document}

\begin{titlepage}
%
%
\vspace{20mm}
\begin{center}
\bf\large
The Role of Vortex Strings in the\\
Non-Compact Lattice Abelian Higgs Model
\end{center}
\vspace{5mm}
\begin{center}
{\bf Michael Chavel}\\
Department of Physics, University of Illinois at Urbana,\\
1110 W. Green St, Urbana, IL 61801-3080.
\end{center}
%
%
\vspace{10mm}     
\begin{abstract}
The non-compact lattice version of the Abelian Higgs model is studied in
terms of its topological excitations.  The Villain form of
the partition function is represented as a sum over world-sheets of
gauge-invariant ``vortex'' strings.  The phase transition of the system is then
related to the density of these excitations.  Through Monte Carlo
simulations the density of the vortex sheets is shown to be a good
order parameter for the system.  The vortex density essentially vanishes
in the Higgs phase, and the Coulomb phase consists of a single
vortex condensate. 
\end{abstract}
\vfill
%
%
\end{titlepage}

%
%
%
%

\section{Introduction}
In many lattice models the use of topological excitations is 
helpful in understanding the phases of the system.  In compact
$\mbox{U}(1)$ gauge theory, for example, the confining transition is
driven by the condensation of magnetic monopoles \cite{BMK,DT,BS}.  These
monopoles are consequences of the periodicity in the lattice gauge
action.  Monopoles also appear to be responsible for the
confinement-deconfinement transition in the compact Abelian Higgs
model.  However, they do not explain the transition separating
the Higgs phase from the Coulomb and confinement phases
\cite{ES,RKR,LSVW}.  This transition appears to be associated with
vortex-like excitations, due to the $\mbox{U}(1)$ symmetry in the
Higgs part of the action \cite{ES,RKR,AP}.
In this paper we will study the role of these excitations in the
\emph{non-compact} version of the Abelian Higgs model.  Because the gauge
action is not compact, no monopoles appear in the theory.\footnote{
%
One can continue to define monopoles exactly as in the compact case.
However they do not enter into the action and have no bearing on the
phases of the system.  For a study of this see ref. \cite{BFKK}.} 
Therefore, we are able to show more clearly how the vortex excitations
influence the Coulomb-Higgs transition.    

The non-compact model is defined as
\be
\label{zorig}
Z &=& \int_{-\infty}^{\infty}\! \int_{-\infty}^{\infty}\! \D\phi \D A_{\mu}
\,\e^{-S} \qquad  S = S_g + S_h \\
\nonumber \\ 
S_g \! &=& \! \textstyle{\beta \over 4}\displaystyle{\sum_{x,\mu\nu}}F_{\mu\nu}^2(x) 
\qquad \qquad \quad \ 
F_{\mu\nu}(x)=\Delta_{\mu}A_{\nu}(x)-\Delta_{\nu}A_{\mu}(x) \nonumber \\
S_h \! &=& \! -\gamma\sum_{x,\mu}^{}(\phi_x^*U_{x,\mu}\phi_{x+\mu} + c.c ) 
+ \sum_x^{}\phi_x^*\phi_x + \lambda \sum_x^{}(\phi_x^*\phi_x - 1)^2 \nonumber
\ee
$\phi_x=\rho(x)\e^{\i\chi(x)}$ is the complex scalar Higgs field, and
$ U_{x,\mu}= \e^{\i A_{\mu}(x)}$.    
$S_h$ is the same as in the compact model. 

\begin{figure}
\begin{center}
\leavevmode
\epsfxsize=3.5in
\epsfbox{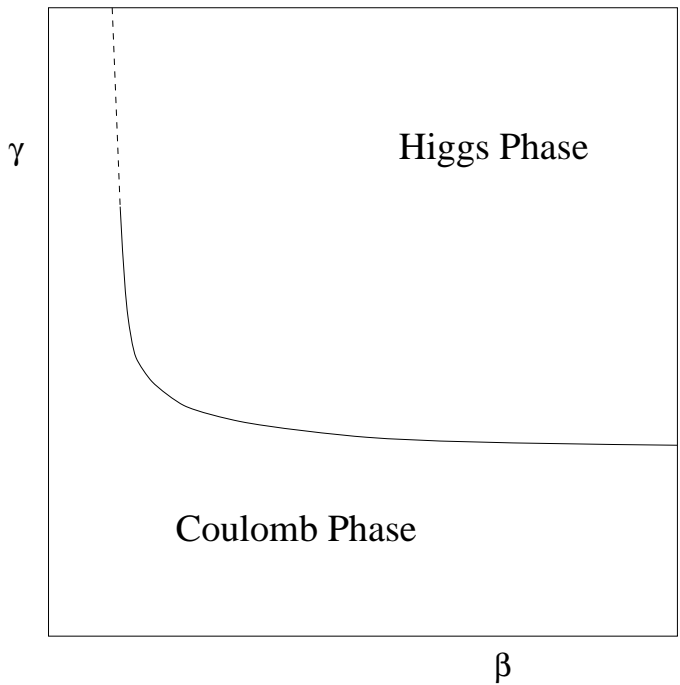}
\caption{The phase structure of the non-compact Abelian Higgs model, for
$\lambda \to \infty$. \label{figphase}}
\end{center}
\end{figure}

It has become customary in both the compact and non-compact cases
to consider $S_h$ in the limit where the Higgs self
coupling diverges $(\lambda \to \infty)$.  In this limit the
magnitude of the Higgs field is constrained to unity, leaving only an
angular degree of freedom.  The resulting total action is 
\be
\label{sfl}
S = \textstyle{\beta \over 4}\displaystyle{\sum_{x,\mu\nu}}F_{\mu\nu}^2(x) 
-2\gamma \sum_{x,\mu}^{} \cos(\Delta_{\mu}\chi(x)-A_{\mu}(x)).
\ee
This fixed-length model is believed to lie in the same universality
class, and it is simpler to analyze numerically.  
The phases of the system have already been determined in previous Monte Carlo
studies \cite{SQED,BFKK}.  A sketch of the phase diagram is shown in
Fig. \ref{figphase}. 
Large values of $\beta$ and $\gamma$ correspond
to a Higgs phase with a massive photon.  While small values of $\beta$
or $\gamma$ correspond to a Coulomb phase, with a massless photon. 
As $\beta \to \infty$ the model reduces to the XY model, which has a
continuous transition at $\gamma_{\mathrm{c}}=0.15$.
The $\gamma \to \infty$ limit describes an integer Gaussian model (the
``frozen superconductor'' of ref. \cite{P}), with
a first order transition at $\beta_{\mathrm{c}}=0.04$. 
See ref. \cite{BN} for a review of analytical results.

Now consider the role of the scalar field in the action. 
The action must be gauge-invariant.
Hence, $S_h$ depends upon the phase of the scalar field only
through the gauge invariant quantity $\Delta_{\mu}\chi(x)-A_{\mu}(x)$.
Since $\chi(x)$ is an angular variable it can be
multi-valued along a ``vortex'' string (the lattice analog of a
Nielsen-Olesen string \cite{NO}), and cannot
be gauged away.  The presence of these strings in the theory is very
important.  Without them the fixed-length model would, necessarily,
describe just a free gauge boson in the continuum limit.  Therefore, these
excitations should be central to determining the phases of the system at large
length scales.  

Einhorn and Savit \cite{ES} have shown how 
to express the partition function of the \emph{compact} model in its dual
form, consisting of sums over closed vortex strings and open vortex
strings terminating on monopoles.  Polikarpov, Wiese, and Zubkov
\cite{POLIKARPOV} have found a similar expression using the Villain
form of the non-compact model.  In the non-compact case only closed vortex
strings appear due to the lack of periodicity in $S_g$.
In the following section we will argue for the existence of a phase
transition driven by the condensation of these strings, which can be
identified with the Coulomb-Higgs transition.  The intuitive
explanation being that the vortices act to disorder the scalar field.
(As $\beta \to \infty$ this is consistent with the results of ref.
\cite{BWPP} in four dimensional XY model.)  In section 3 the
results of numerical simulations will be presented demonstrating that
this picture is indeed correct.

\section{The Villain Approximation}

The Villain form \cite{V} of the fixed-length partition function is
\be
\label{zvill}
Z= \int_{-\infty}^{\infty}\! \int_{-\infty}^{\infty}\! \D\chi \D A_{\mu}
\sum_{a_{\mu}=-\infty}^\infty
\e^{-{\beta \over 4}\sum F_{\mu\nu}^2
-\gamma\sum(\Delta_{\mu}\chi-A_{\mu}+2\pi a_{\mu})^2}
\ee
(where we have suppressed the spatial dependence of all the fields).
$a_{\mu}$ is an integer-valued field which preserves the periodic
symmetry of the original action.  This form can be thought of as a
``low-temperature'' (large $\gamma$) Gaussian
approximation to the action around each minimum of $S_h$. 

By a series of exact transformations, the partition
function above can be rewritten as 
\be
\label{zvtx}
Z \sim \sum_{a_{\mu}=-\infty}^\infty \e^{-4\pi^2\gamma 
\sum J_{\mu\nu}(x) D(x-y\,;\, m^2) J_{\mu\nu}(y)}
\ee
where $J_{\mu\nu} = \epsilon_{\mu\nu\lambda\rho}\Delta_{\lambda}a_{\rho}$
is an integer-valued ``vortex'' current.  $D(x-y;\ m^2)$ is the usual
lattice Green's function, with $m^2=\textstyle{2\gamma\over\beta}$.
The details of the derivation are provided in the appendix.
The identification of $J_{\mu\nu}$ as the vorticity in the integer part
of the Higgs action should be clear.
Notice that $\Delta_{\mu}J_{\mu\nu}=0$. Consequently, $J_{\mu\nu}$
forms closed sheets (world-sheets of closed strings) on the dual lattice.

Now consider the possible phases of (\ref{zvtx}) in terms of
the vortex excitations.  Due to the behavior of the lattice
Green's function, the self energy of the vortex sheets will be much larger
than any spatially separated interactions.  The
partition function can then be approximated by retaining only the
diagonal terms. 
\be
Z \sim \sum_{a_{\mu}=-\infty}^\infty \e^{-4\pi^2\gamma D(0\,;\, m^2) 
\sum J_{\mu\nu}^2(x)}
\ee
For reasonably large $\gamma$, configurations with
$J_{\mu\nu}(x) = 0, +1,-1$ will be dominant.
The energy of a vortex sheet of area A will be $E_A =
4\pi^2\gamma D(0\,;\,m^2)A$. Now consider the entropy of such a configuration.
Let $N_A$ be the number of possible closed sheets of area $A$ containing
a given plaquette of the lattice.  For reasonably large $A$ the
leading behavior will be $N_A \sim \mu^A$.\footnote{
%
See the second paper of ref. \cite{ES} for a derivation of this result.
$\mu > 1$.  However, a good estimate of $\mu$ is unknown to the author.}
The entropy of a sheet of size $A$ will then be $S_A \sim A\log\mu$. 
Thus, balancing energy and entropy produces the critical line 
\be
\label{cl}
\log\mu = 4\pi^2 \gamma D(0 \, ; \textstyle{2\gamma\over\beta}),
\ee  
where sheets of all sizes are expected to occur.\footnote{
%
The same critical line exists in the Villain approximation ($\gamma,
\beta \gg 1$) to the compact model \cite{ES}.  The two models should
agree at large values of $\beta$.  In this case, however, no
approximation has been made to $S_g$ so there is no restriction on the
values of $\beta$.}

By considering the behavior of $D(0\,;\,m^2)$ it can be seen that the
critical line above is consistent with the phase diagram of Fig \ref{figphase}.
For example, as $\gamma \to \infty \ D(0\,;\,m^2) \sim {1 \over m^2}$,
producing a critical point at
\be
\beta_{\mathrm{c}} \approx {\log\mu \over 2\pi^2}.
\ee
For larger values of $\beta$ only relatively small surfaces will be
present, leading to an ordered (Higgs) phase.  For smaller $\beta$
values the surfaces are allowed to grow large and intersect, forming a
single vortex condensate and a disordered (Coulomb) phase. The
Coulomb-Higgs transition can then be interpreted as  a condensation of
these vortex sheets.

\section{Numerical Results}

Now let's reconsider the original form of the partition function
(\ref{zorig}). The conclusions of the previous section can be motivated here 
by examining the periodicity of the lattice Higgs action.  For each
link of the lattice perform the decomposition
\be
\Delta_{\mu}\chi(x)-A_{\mu}(x) \equiv \theta_{\mu}(x)+2\pi a_{\mu}(x),
\ee
where $\theta_{\mu}(x) \in (-\pi,\pi)$ and $a_{\mu}(x)$ is an integer.
Notice that changing $a_{\mu}(x)$ on any link of the
lattice leaves $S_h$ unchanged.  The result is the creation of a Dirac
string (or vortex) threading through each of the plaquettes of which
the link is an element.  The string current being the same as that defined
previously\footnote{
%
The same definition was used in studying the compact model in
ref. \cite{RKR}.  Notice the similarity with the definition of
monopoles \cite{DT} in compact QED.  In compact gauge theories the
gauge action is periodic in the plaquette variables.  The monopoles
are then taken to be the oriented sum of the integer parts of the
plaquettes around each 3-cube.}
\be
\label{vcurrent}
J_{\mu\nu}(x)=\epsilon_{\mu\nu\lambda\rho}\Delta_{\lambda}a_{\rho}(x).
\ee
If $\gamma$ is reasonably large the fluctuations in $\theta(x)$ will be
small.  The entropy of the system will be entirely due to
$J_{\mu\nu}(x)$. The energy being $S_g=2\pi^2\beta\sum J^2$.
Thus, we arrive at the same critical energy-entropy balance we found
in the Villain approximation.

It is not clear if this can be carried
over to smaller values of $\gamma$ where the fluctuations in $\theta(x)$
are not small. To examine the behavior of the vortex excitations at
these smaller values of $\gamma$ Monte Carlo simulations of the fixed-length
model were performed.  The simulations were done on a $8^4$ lattice with
periodic boundary conditions, and in the unitary gauge.  The vortex
currents were then measured directly, using the definition above.

\begin{figure}
\begin{center}
\leavevmode
\epsfxsize=4in
\epsfbox{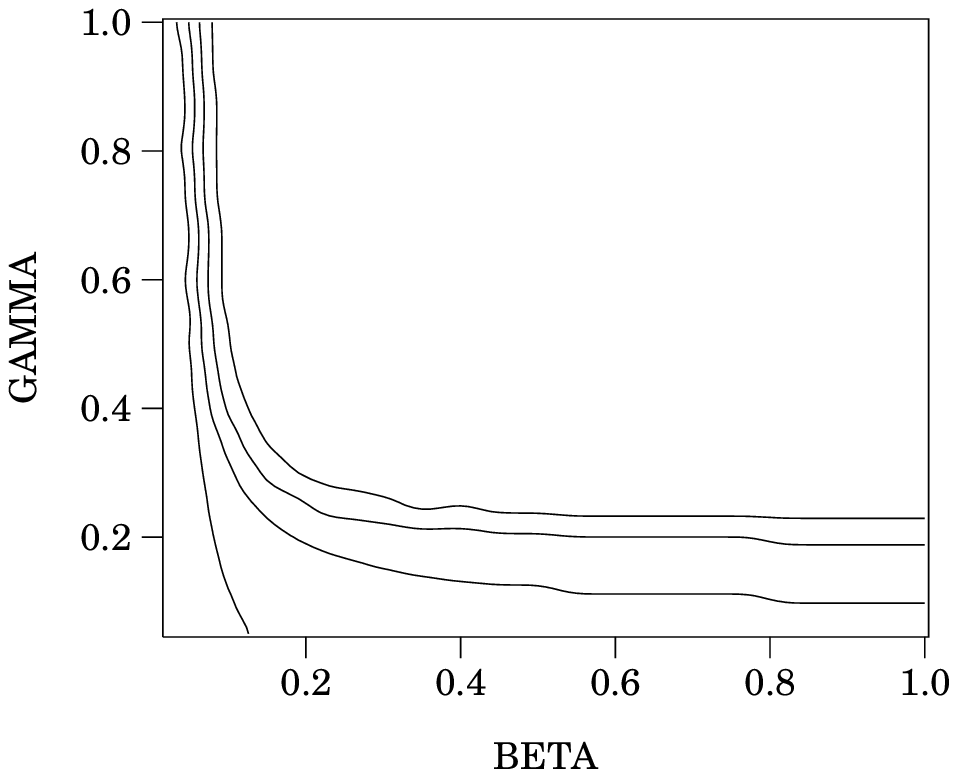}
\caption{$V$ plotted as lines of constant vortex density.
$V=0.4,0.3, 0.2, 0.1$, from left to right. \label{figvc}} 
\leavevmode
\epsfxsize=3.5in
\epsfbox{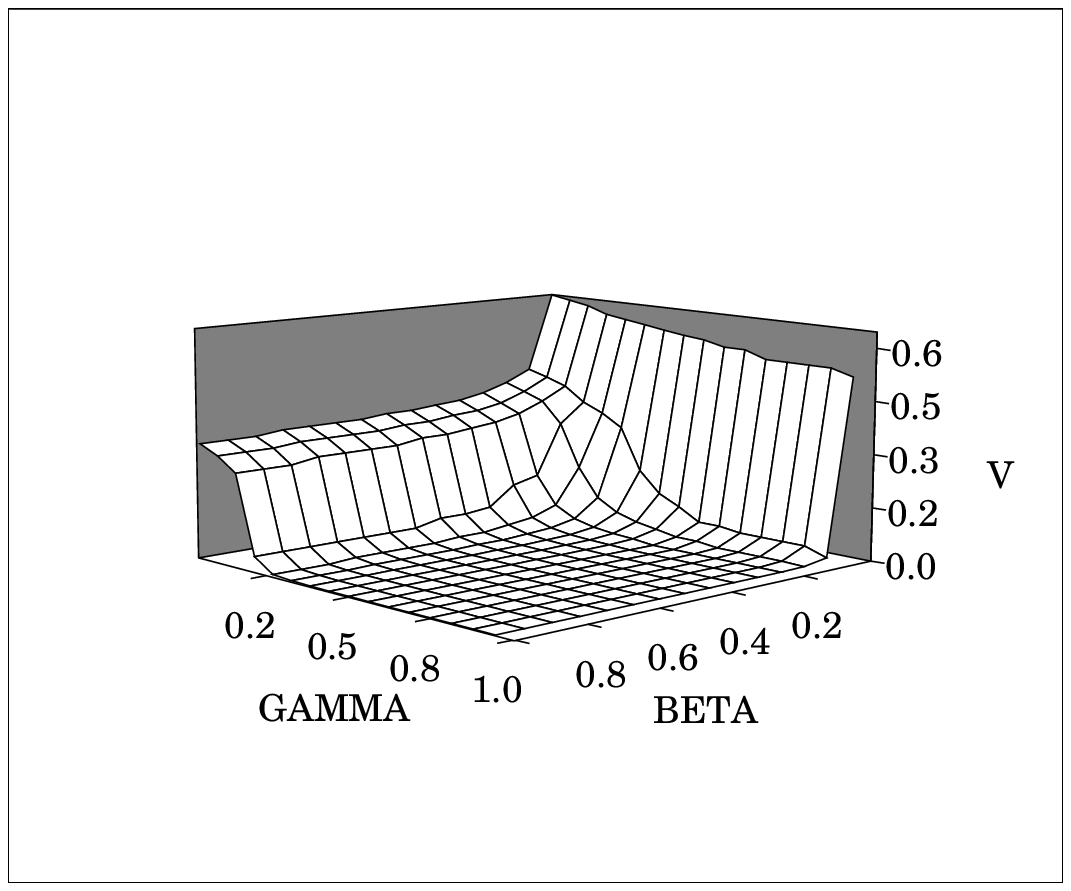}
\caption{Surface plot of the vortex density, $V$, as measured on a
$8^4$ lattice. \label{figvs}}
\end{center}
\end{figure}

\begin{figure}
\begin{center}
\leavevmode
\epsfxsize=5.4in
\epsfbox{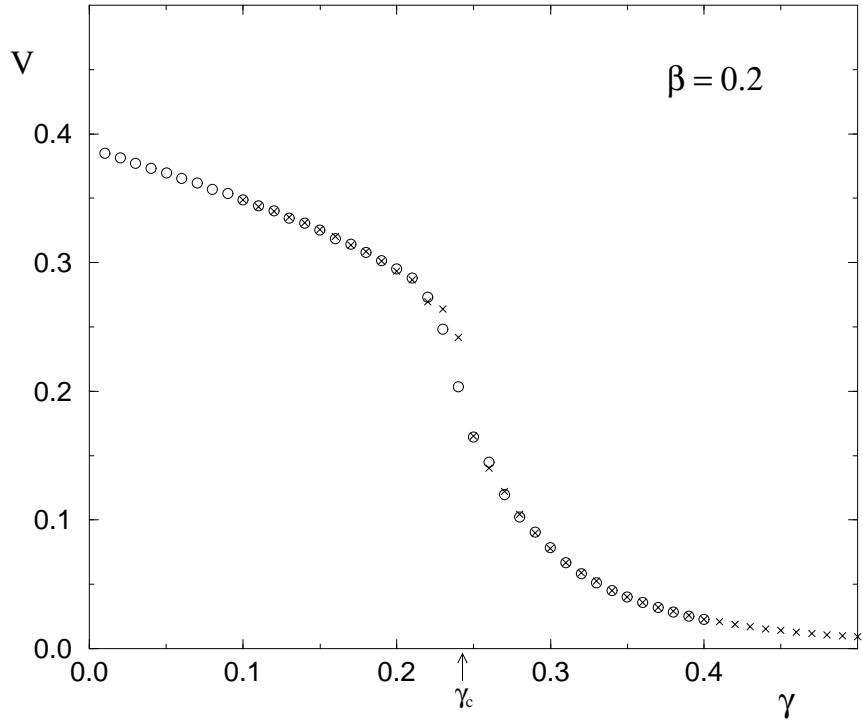}
\caption{Measurements of $V$ on a $8^4$ lattice, at $\beta = 0.2$.
Crosses correspond to data taken with increasing $\gamma$ steps.
Circles correspond to decreasing $\gamma$ steps. \label{figvbeta0.2}}
\end{center}
\end{figure}

The simplest measurement which
can be made is the amount of vortex current per dual plaquette
\be
\label{vdnsty}
V \equiv \biggl<\,{1 \over N_p} \sum_{x,\mu > \nu} |J_{\mu\nu}(x)|\, \biggr>,
\ee
where $N_p$ is the number of plaquettes on the lattice.
Figures \ref{figvc} and \ref{figvs} show the results of such
measurements for $\gamma$ and $\beta$ ranging from $0$ to $1$.
$V$ essentially vanishes in the Higgs phase and
rises sharply across the phase boundary to a non-zero value in the
Coulomb phase.  Figure \ref{figvbeta0.2}  shows separate ``heating''
and ``cooling'' runs at fixed $\beta = 0.2$.
The phase transition, given by the peak in the specific heat, is
marked by the arrow at $\gamma_{\mathrm{c}}=0.24$. Statistical errors
are of the size of the data points.  The hysteresis in the data at the
critical point appears indicative of the continuous (2nd order) phase
transition there.  Runs at other values of the couplings produced
similar, excellent, agreement.  Therefore, $V$ appears to be a good
order parameter for the system.      

To better understand the distribution of the vortices,
a routine was written to measure the area of each connected vortex
sheet.  The results support the description of the phase transition
given in section 2.  Deep into the Higgs phase only small sheets are
present, containing six or ten dual plaquettes each.  As the system is
moved towards the transition the average size of the clusters increases.
Near the transition the clusters begin to coalesce into one large cluster, and
the number of separate clusters drops dramatically.  This occurs in
the Higgs phase, before the phase boundary is reached.  When the system
reaches the Coulomb phase there is only a single large cluster, roughly
the size of the lattice.  If the system is heated further, the density
of this cluster continues to increase. 
    
In conclusion, the numerical simulations presented here seem to
confirm the picture presented by the large $\gamma$ analysis of the
theory.  In the Higgs phase the vortex density is low and the area of
each sheet is relatively small.  In the Coulomb phase the vortex
sheets are large and overlap, forming a single cluster of dual
plaquettes.  The abrupt change in the vortex density is coincident
with the transition, given by the peak in the specific heat.  Hence,
the phase transition in the non-compact Abelian Higgs model, for
$\lambda \to \infty$, appears to be driven by the condensation of
vortex excitations.

The author would like to thank John B. Kogut for assistance with the Monte
Carlo simulations and other helpful discussions.  The author was
supported in part by a GAANN fellowship.

\section*{Appendix}
\setcounter{section}{1}
\setcounter{equation}{0}
\def\theequation{\Alph{section}.\arabic{equation}}

Here it is shown how to transform the Villain form of the
partition function into equation (\ref{zvtx}).  The steps
are essentially the same as those used in analyzing the compact
model \cite{ES}.  For greater clarity we will suppress the spatial
dependence of the fields.  (See ref. \cite{POLIKARPOV} for a
derivation using lattice differential forms.) 

We start with
\be
Z= \sum_{a_{\mu}=-\infty}^{\infty}\! 
\int_{-\infty}^{\infty} \D\chi \D A_{\mu}
\,\e^{-{\beta \over 2}\sum (\epsilon_{\mu\nu\lambda\rho}
\Delta_{\lambda}A_{\rho})^2
-\gamma\sum(\Delta_{\mu}\chi-A_{\mu}+2\pi a_{\mu})^2}
\ee
and use the identity
\be
\e^{-\textstyle{\kappa \over 2}x^2} = \sqrt{1 \over {2\pi\kappa}}
\int_{-\infty}^{\infty}\! \d y\,\e^{-{y^2 \over {2\kappa}}+\i y x}
\ee
to rewrite each of the Gaussian integrals.
\be
Z &\sim& \sum_{a_{\mu}=-\infty}^\infty \int_{-\infty}^{\infty}
\D\chi \D A_{\mu} \D L_{\mu} \D S_{\rho\sigma}\nonumber\\
 & &\mbox{ }\e^{\sum \bigl[-{S_{\rho\sigma}^2 \over {2\beta}}+
\i S_{\rho\sigma}\epsilon_{\rho\sigma\mu\nu}\Delta_{\mu}A_{\nu}
-{L_{\mu}^2 \over {4\gamma}}+
\i L_{\mu}(\Delta_{\mu}\chi-A_{\mu}+2\pi a_{\mu})\bigr]}.
\ee
Integrating over $\chi$ and $A_{\mu}$ gives
\be
Z &\sim& \sum_{a_{\mu}=-\infty}^\infty
\int_{-\infty}^{\infty} \D L_{\mu} \D S_{\rho\sigma}
\delta(\Delta_{\mu}L_{\mu})
\delta(L_{\mu}+\epsilon_{\mu\nu\rho\sigma}\Delta_{\nu}S_{\rho\sigma})
\nonumber\\
& &\mbox{ }\e^{\sum \bigl[-{S_{\rho\sigma}^2 \over {2\beta}}
-{L_{\mu}^2 \over {4\gamma}}
+2\pi \i L_{\mu}a_{\mu}\bigr]}.
\ee
The constraint imposed by $\delta(\Delta_{\mu}L_{\mu})$ can be removed
by letting
\be
L_{\mu}=\epsilon_{\mu\nu\rho\sigma}\Delta_{\nu}B_{\rho\sigma}.
\ee
The second constraint then becomes
\be
\epsilon_{\mu\nu\rho\sigma}\Delta_{\nu}(B_{\rho\sigma} +
S_{\rho\sigma}) = 0,
\ee
implying that
\be
B_{\rho\sigma} + S_{\rho\sigma} =
 \Delta_{\rho}\psi_{\sigma}-\Delta_{\sigma}\psi_{\rho}.
\ee
We still must choose a gauge for $B_{\rho\sigma}$ to insure that $L_{\mu}$ is
properly defined.  The most convenient choice is to take 
$B_{\rho\sigma} + S_{\rho\sigma} = 0$. The partition function is then
\be
\label{zabove}
Z &\sim& \sum_{a_{\mu}=-\infty}^\infty \int_{-\infty}^{\infty}
\D B_{\rho\sigma}\nonumber \\
& & \mbox{ }\e^{-\textstyle{1 \over{4\gamma}}\sum
\bigl[(\epsilon_{\mu\nu\rho\sigma}\Delta_{\nu}B_{\rho\sigma})^2
+ m^2 B_{\rho\sigma}^2 +2\pi\i a_{\mu}
(\epsilon_{\mu\nu\rho\sigma}\Delta_{\nu}B_{\rho\sigma})\bigr]},
\ee
with $m^2 \equiv \textstyle{2\gamma \over \beta}$.
Finally, integrating over $B_{\rho\sigma}$, we get the desired result
\be
Z \sim Z_o\! \sum_{a_{\mu}=-\infty}^\infty
\e^{-4\pi^2\gamma \sum J_{\mu\nu}(x)D(x-y;m^2)J_{\mu\nu}(y)}, 
\ee
where $(-\Delta^2 + m^2)D(x-y;m^2)=\delta_{xy}$ and 
$J_{\mu\nu} \equiv
\epsilon_{\mu\nu\lambda\sigma}\Delta_{\lambda}a_{\sigma}$.  
$Z_o$ is the determinant given by setting all the $a_{\mu}=0$ in
(\ref{zabove}), and is completely analytic.

\vfill


\vfill
\newpage

\begin{thebibliography}{9}

\bibitem{BMK}
T. Banks, R. Myerson, and J. Kogut,
Nucl.~Phys.~{\bf B129} (1977) 493.

\bibitem{DT}
T. A. DeGrand and D. Toussaint,
Phys. ~Rev.~{\bf D 22} (1980) 2478.

\bibitem{BS}
J. S. Barber and R. E. Shrock,
Nucl.~Phys.~{\bf B257} (1985) 515.

\bibitem{ES}
M. B. Einhorn and R. Savit,
Phys.~Rev.~{\bf D 17} (1978) 2583;
Phys.~Rev.~{\bf D 19} (1979) 1198.

\bibitem{RKR}
J. Ranft, J. Kripfganz, and G. Ranft,
Phys.~Rev.~{\bf D 28} (1983) 360.

\bibitem{LSVW}
J.M.F. Labastida, E. Sanchez-Velasco, R. E. Shrock and P. Wills,
Nucl.~Phys.~{\bf B 277} (1986) 393.

\bibitem{AP}
M. G. Amaral and M. E. Pol,
Z.~Phys.~{\bf C 44} (1989) 515.

\bibitem{SQED}
D. Callaway and R. Pertronzio,
Nucl.~Phys.~{\bf B 277} (1986) 50;
J. Jersak,
in: Lattice Gauge Theory - A Challenge in Large-Scale Computing,
eds. B. Bunk, K.H. M\"utter and K. Schilling (Plenum Press, 1986)
p.133;
M. Baig, E. Dagotto, J. Kogut, and A. Moreo
Phys.~Lett.~{\bf B 242} (1990) 444.

\bibitem{BFKK}
M. Baig, H. Fort, J. Kogut, and S. Kim,
Phys.~Rev.~{\bf D 51} (1994) 5216. 

\bibitem{P}
M. E. Peskin,
Ann.~Phys.~{\bf 113} (1978) 122.

\bibitem{BN}
C. Borgs F. Nill,
J.~Stat.~Phys~{\bf 47} (1987) 877.

\bibitem{NO}
H. B. Nielsen and P. Olesen,
Nucl.~Phys.~{\bf B61} (1973) 45.

\bibitem{POLIKARPOV}
M.I. Polikarpov, U.-J. Wiese, M.A. Zubkov,
Phys.~Lett.~{\bf B 309} (1993) 133;
M. A. Zubkov and M. I. Polikarpov, 
JETP Lett. {\bf 57} (1993) 461.

\bibitem{BWPP}
A. K. Bukenov, U.-J. Wiese, M. I. Polikarpov, and Pochinskii,
Phys.~At.~{\bf 56} (1993) 122.

\bibitem{V}
J. Villain,
J.~Phys.~(Paris)~{\bf 36} (1975) 581.

\end{thebibliography}
\end{document}